\documentclass{article}

\usepackage{arxiv}

\usepackage[utf8]{inputenc} 
\usepackage[T1]{fontenc}    
\usepackage{hyperref}       
\usepackage{nicefrac}       
\usepackage{microtype}      
\usepackage{lipsum}
\usepackage{graphicx}
\usepackage{subfigure} 
\usepackage{color}
\usepackage[english]{babel} 
\usepackage{amsmath,amsfonts,amsthm} 
\usepackage{float}
\usepackage{algorithmic}
\usepackage{enumerate} 
\usepackage{array}
\usepackage{multirow}
\usepackage{xfrac}
\usepackage[table]{xcolor}
\usepackage{lineno}
\definecolor{lg}{gray}{0.9}

\usepackage{url}

\title{Classifying sleep-wake stages through recurrent neural networks using pulse oximetry signals}

\author{
  Ramiro Casal \\
  Lab. Señales y Dinánicas no Lineales\\
  IBB - UNER - CONICET \\
  Oro Verde (3100), Entre Ríos, Argentina  \\
  \texttt{rcasal@conicet.gov.ar} \\
  \And
  Leandro E. Di Persia \\
  sinc(i)\\
  FICH - UNL - CONICET \\
  Santa Fe (3000), Santa Fe, Argentina \\
  \texttt{ldipersia@sinc.unl.edu.ar} \\
  \AND 
  Gastón Schlotthauer \\
  Lab. Señales y Dinánicas no Lineales\\
  IBB - UNER - CONICET \\
  Oro Verde (3100), Entre Ríos, Argentina  \\
  \texttt{gschlotthauer@ingenieria.uner.edu.ar} \\
}

\begin{document}
\maketitle

\begin{abstract}
The regulation of the autonomic nervous system changes with the sleep stages causing variations in the physiological variables. We exploit these changes with the aim of classifying the sleep stages in awake or asleep using pulse oximeter signals. We applied a recurrent neural network to heart rate and peripheral oxygen saturation signals to classify the sleep stage every $30$ seconds. The network architecture consists of two stacked layers of bidirectional gated recurrent units (GRUs) and a softmax layer to classify the output. In this paper, we used $5000$ patients from the Sleep Heart Health Study dataset. $2500$ patients were used to train the network, and two subsets of $1250$ were used to validate and test the trained models. In the test stage, the best result obtained was $90.13\%$ accuracy, $94.13\%$ sensitivity, $80.26\%$ specificity, $92.05\%$ precision, and $84.68\%$ negative predictive value. Further, the Cohen's Kappa coefficient was $0.74$ and the average absolute error percentage to the actual sleep time was $8.9\%$. The performance of the proposed network is comparable with the state-of-the-art algorithms when they use much more informative signals (except those with EEG).\end{abstract}


\section{Introduction} \label{intro}

Sleep studies are important to evaluate sleep and sleep-related pathologies. The gold standard test for diagnosing sleep disorders is a polysomnography study (PSG), during which several physiological signals are recorded simultaneously in a specially conditioned sleep medical center. These recordings include signals as electroencephalography (EEG), electrocardiography (ECG), electromiography (EMG), respiratory effort and oronasal airflow, peripheral oxygen saturation ($\text{SpO}_2$) and heart rate (HR), among others. The PSG study needs to be supervised by a technician and its analysis requires a tedious manual scoring, usually done with the help of a software. For this reason, PSG is an expensive and scarcely available study. Further, the patients frequently have trouble falling asleep, so the studies needs to be repeated \cite{pang2006screening}. Also, scoring has been reported to suffer from high inter-professional variability \cite{norman2000interobserver}. 

Due to these disadvantages, many studies have been proposed with the aim of developing methods of diagnosis alternatives to PSG. The use of screening methods for sleep disorders would reduce the need for PSG studies in cases where it is not strictly necessary. Cardiac and respiratory sounds \cite{yadollahi2006apnea}, electrocardiography (ECG) \cite{roche2004heart}, nasal airway pressure \cite{salisbury2007rapid}, pulse oximeter \cite{schlotthauer2014screening, hang2015validation, rolon2017discriminative} and combinations of various signals \cite{raymond2003combined} have been proposed for screening sleep pathologies. Pulse oximeter is an ideal choice for the screening due to its low cost, accessibility and simplicity \cite{pang2006screening}.

One of the most prevalent sleep disorders diagnosed by PSG is obstructive sleep apnea/hypopnea syndrome (OSAHS)\cite{sateia2014international}. OSAHS is characterized by repetitive upper airway obstructions producing partial or total reduction in the airflow. The most important index to evaluate OSAHS severity is the apnea/hypopnea index (AHI), which represents the number of apnea/hypopnea events per hour of sleep. 

Several studies attempt to estimate AHI from pulse oximeter signals \cite{schlotthauer2014screening, hang2015validation, rolon2017discriminative}. AHI can be approximated by oxygen desaturation index (ODI), which is estimated by counting the blood oxygen desaturations per hour of sleep. These desaturations are related with apnea and hypopnea events. However, the aforementioned works do not take into account whether the patient is sleep or not by using pulse oximeter signals. In some cases, the total sleep time ($\mathrm{TST}$) is approximated by the total recording study ($\mathrm{TRT}$), resulting in an underestimation of the AHI \cite{corral2017conventional}. Sabil et al. mentioned the consequences of the overestimation of $\mathrm{TST}$ in home studies to diagnose sleep apnea, and how its influence in the AHI can lead to underestimation of the index, resulting in underdiagnosis \cite{sabil2018automatic}. In other cases, $\mathrm{TST}$ is estimated using the EEG, even when it would not be available in a real at-home sleep test. 

The aforementioned drawbacks can be overcome by developing methods to estimate $\mathrm{TST}$ from signals obtained by screening devices. Motivated by this, the aim of our work is to classify the sleep stage in awake (W) or asleep (S) using signals from pulse oximeter, namely HR estimated from photoplethysmography (PPG) and  $\text{SpO}_2$. For this reason, this work is related to the automatic sleep staging problem. In this field of application, the cutting edge of performance is obtained from EEG \cite{fraiwan2012automated}. Nevertheless, there are many researchers whose objective is to develop algorithms for automatic sleep staging using more portable screening devices, which do not include EEG recordings. 

The dynamic of HR variability changes with sleep stages \cite{penzel2003dynamics, aeschbacher2016heart}. Based on this relationship, several works developed algorithms to classify sleep stages from cardiac-related signals as ECG \cite{adnane2012sleep, yucelbacs2018automatic, malik2018sleep} and photoplethysmography (PPG) from pulse oximeter \cite{uccar2016automatic, casal2019sleep, beattie2017estimation}. The best performance so far is obtained by Beattie et al.\cite{beattie2017estimation}, but in that work the authors used a complementary accelerometer signal. Most of these works have used classical machine learning approaches to classify the sleep stages. They extracted and selected several features from the signals, which are then used as input of the classifier. Conversely, Malik et al. \cite{malik2018sleep} have used convolutional neural networks. In our previous work we also have used a classical method for classification \cite{casal2019sleep}. But, unlike others works, we have used a large database that allows to determine the generalization capability of the developed algorithms.

In this work, classification in awake/sleep will be done by applying recurrent neural networks (RNNs) to HR and $\text{SpO}_2$ signals, with a very simple preprocessing step. The RNNs are able to process and classify the pulse oximeter signals by taking advantage of the information about the entire sequence stored in the ``state vectors'' to learn the temporal dependencies of the internal structure of the sleep \cite{lecun2015deep}. In contrast to Malik et al. \cite{malik2018sleep}, we use a simpler network architecture, but achieving a result that is comparable to the one of Beattie et al.  \cite{beattie2017estimation} without using any complementary signal.

The accurate estimation of $\mathrm{TST}$ using a pulse oximeter device would improve the detection and characterization of OSAHS. In addition, awake/asleep classification may be useful for many other applications. For example, automatic systems with the goal of detecting and preventing drowsy drivers from falling asleep are an active area of research. Most of those systems use video cameras, to assess sleepiness by detection of physiological events related to fatigue and drowsiness \cite{dong2011driver}. Due to its characteristics, our algorithm could provide complementary information on these systems. Furthermore, daily life applications related with sleep measures from personal health monitoring devices are currently under spotlight \cite{mantua2016reliability}. In summary, any critical work in which the sleepiness can cause accidents and material or human losses can benefit from applications such as the one developed in this paper.

The principal contributions of our work are as follows:

\begin{itemize}
 \item We present an RNN-based architecture to perform a classification, without using hand-engineered features or any auxiliary signal. The RNN can be trained to learn the temporal dependencies of the sleep stages. Despite using simple-to-acquire signals, the obtained performance is comparable to the state of the art.
 \item We assess architectures with different parameters and input signals in a large database in order to achieve the optimal network  to permit a fast screening of sleep staging. Further, the developed algorithm is useful to be adapted for apnea screening methods and other applications like drowsy driver monitoring and wearable devices for personal health monitoring.
\end{itemize}

The layout of the article is as follows. In section \ref{sec:materials} we formally define and describe sleep stages, explaining how heart-related signals are affected by them and how these changes can be shown by the pulse oximeter signals. Further, we present and describe the used database. In section \ref{sec:methods} we concisely explain the architecture of the designed network and the principal concepts related to RNNs. In section \ref{sec:results} we show the results achieved with this algorithm and present all the parameter configurations needed to reproduce these results. Finally, we discuss the use of pulse oximeters to diagnose sleep disorders in section \ref{sec:discussion} and compare our results with the state-of-the-art.

\section{Materials} \label{sec:materials}

\subsection{Sleep stages and oximetry signals}

Typically, the quality of sleep is determined by sleep experts through PSG studies. In these studies, sleep is classified in different sleep stages. 
In clinical practice, there are two available standards that represent a guideline for diagnosing sleep disorders, the traditional Rechtschaffen and Kales (R\&K) \cite{rechtschaffen1968manual} and, since 2007, the more recently standard published by the American Academy of Sleep Medicine (AASM) \cite{berry2012aasm}. According to the R\&K standard, the PSG recordings are split in consecutive $30$ seconds long segments and each segment is classified in \emph{wakefulness} (W), two stages of \emph{light sleep} (N1 and N2), two of \emph{deep sleep} (N3 and N4) and \emph{rapid eye movement sleep} (REM), which are differentiated on the basis of characteristic waveforms that can be found in EEG, EMG and EOG \cite{rechtschaffen1968manual, penzel2003dynamics}. The AASM modifies the R\&K rules with the aim of increasing the inter-rater reliability of sleep staging, unifying N3 and N4 in a single stage, called simply N3 or slow wave sleep. In this work, all the stages corresponding to \emph{asleep}, namely N1, N2, N3 and REM, are considered as a single category.

As mentioned above, several papers have studied the relationship between different sleep stages and HR \cite{penzel2003dynamics, aeschbacher2016heart}. The most common methods for estimating HR are based on ECG \cite{afonso1999ecg}. Nevertheless, there is a relevant interest in developing methods to estimate HR from PPG obtained by pulse oximeter as it is a low cost, simple and portable technology \cite{zhang2014troika}. Pulse oximeter is a medical device that consist of  a light source and a photodetector. This device is used to indirectly screen $\text{SpO}_2$ and detect blood volume changes. Oximetry signals result from light interaction with biological tissues and several variables of clinical interest can be estimated from it \cite{allen2007photoplethysmography}. 

The pulse oximeters provide two signals that are used in this work, namely $\text{SpO}_2$ and HR estimation from the pulsatile component of PPG, which is synchronized to each heartbeat. $\text{SpO}_2$ is estimated using two light sources (red and infrared) that shows absorption differences due to hemoglobin presence \cite{kyriacou2005pulse}. Usually, the algorithms to estimate HR consist of digital filters and zero crossing detectors, especially in commercial devices \cite{allen2007photoplethysmography}. Nonetheless, there are many studies with the aim of developing robust algorithms that are not affected by movement artifacts \cite{zhang2014troika}.


As previously stated, oximetry signals have a relationship with sleep stages. The regulation of autonomic nervous system changes with sleep stages causing variations in many physiological variables. The reduced metabolism during sleep is reflected in a decrease in HR, blood pressure, and respiratory rate. The average HR drops gradually as the sleep stage goes deeper. Instead, REM stage presents greater variability in the HR and a slightly increase in its average \cite{penzel2003dynamics}. The connection between the $\text{SpO}_2$ signal and the sleep is more complex. The apnea events produce a slow decay of oxygen saturation levels and a subsequent fast recovery. Many times these recoveries are associated with an awakening event. These changes of dynamics in heart rate signals and $\text{SpO}_2$ allows us to get information about the sleep stage.

\subsection{\emph{Sleep Heart Health Study} dataset}

In this work we use signals from the \emph{Sleep Heart Health Study} (SHHS) dataset, which was designed to investigate the cardiovascular consequences of OSAHS and other sleep-disordered breathing. Two sets of home PSG records (SHHS 1 and SHHS 2) were obtained from the participants included in the admission criteria. SHHS 1 and 2 have been recorded with a difference of several years in order to study the evolution of the disease in the patients. These database contains several signals corresponding to PSG study acquired automatically at patient's home with supervision of specialized technicians \cite{redline1998methods}. %

The PSG records were processed with a software providing estimations of AHI, arousals, sleep stages, oxygen desaturation events, among others. Then, these outcomes were manually corrected by specialists. Full details about database and signal analysis protocol can be found in \cite{redline1998methods, nieto1997sleep}. In table 1 from Casal et al. \cite{casal2019sleep} can be seen a summary of the principal features of the database related to this work. Further, the average $\mathrm{TST}$ is $587.7 \pm 107.6$.

We used $\text{SpO}_2$ and HR signals, obtained by means of a pulse oximeter, from $5000$ randomly selected patients from SHHS 1. The $\text{SpO}_2$ has a sampling rate of $1$ Hz, resolution of $1\%$ and accuracy of $\pm 2\%$ in the range of $70\%$ to $100\%$. Its performance significantly decreases for values below this range. The HR signal has a sampling rate of $1$ Hz and a precision of $3$ beats per minute. There is an additional quality status signal that provides information about the sensor connection status. This complementary signal consist of two states corresponding to good/defective connection. A good connection is one in which the sensor is correctly in contact with the patient's skin, being able to register the signal with sufficient quality to later be processed.

It is worth clarifying that in this work we only use SHHS 1 to avoid having repeated subjects for the design of the network.

\section{Methods} \label{sec:methods}

\begin{figure}[t]
\centering
\includegraphics[width=0.47\textwidth]{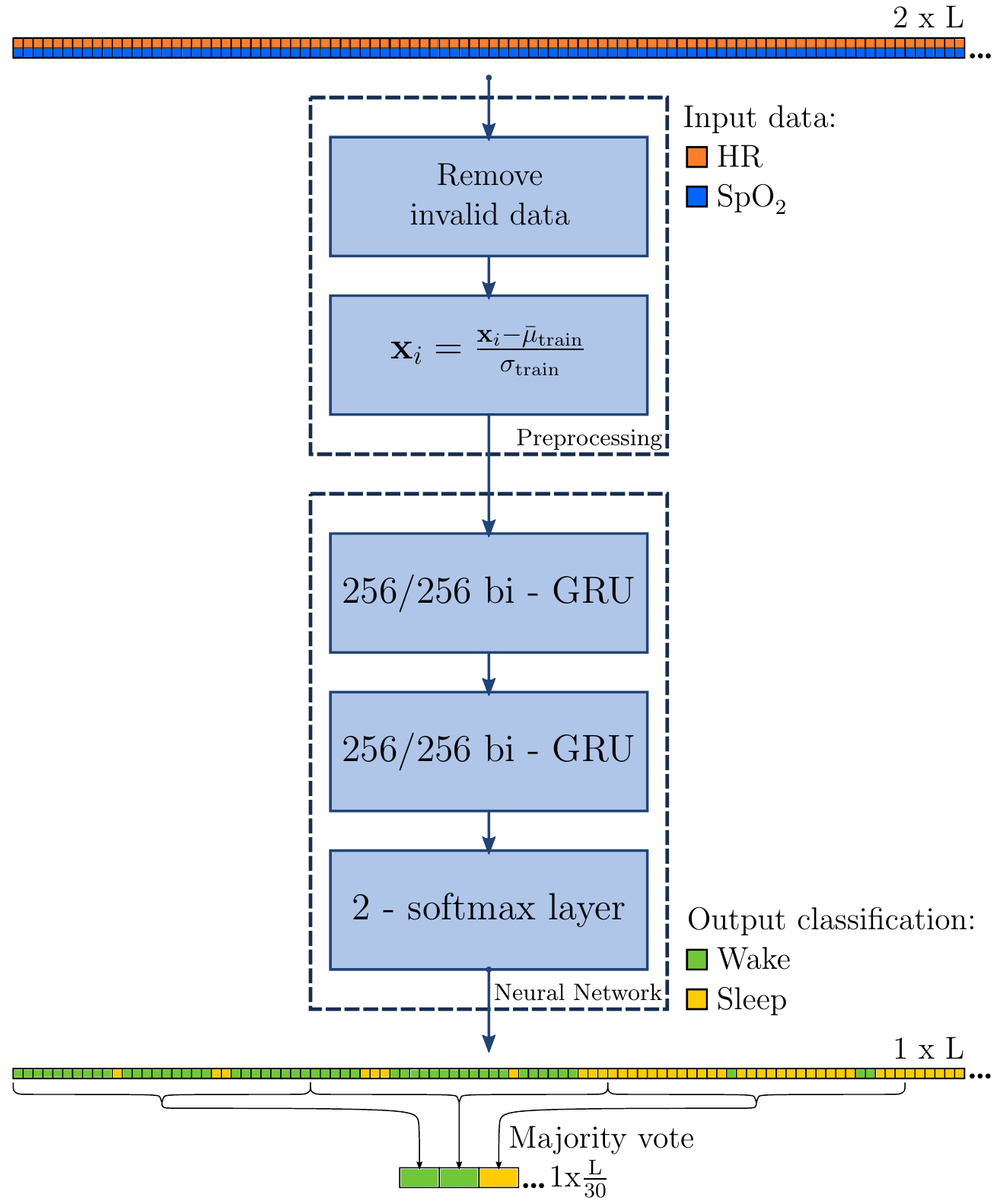}
\caption{An overview of the best architecture consisting of two stacked layers of bidirectional GRU and a softmax layer to classify the outputs of the GRUs.}
\label{fig:network_architecture}
\end{figure}

This paper proposes a deep learning model based on RNN to automatically classify the sleep stage in awake or asleep. We evaluated different architectures of a particular type of RNN called gated recurrent unit (GRU) \cite{cho2014learning}, which is a simplified variant of long short-term memories (LSTM) \cite{hochreiter1997long}. The network architecture consists of two stacked layers of bidirectional GRU. The input data to the network have a simple preprocessing. Due to bidirectionality, the model is able to exploit both past and future information \cite{schuster1997bidirectional}. The outputs of GRUs are classified with a softmax layer. This neural network predict sleep stages for each input. Namely, the length of the input and outputs to the GRU are the same. Then, we performed a majority vote to adapt the classification to the AASM standard. An overview of the best network architecture is shown in figure \ref{fig:network_architecture}.

The algorithm was implemented using Python 3 and Pytorch frameworks and the experiments were run on a cluster of high-performance computing nodes and in a personal computer. 

\subsection{Preprocessing}
The used dataset was randomly split in three subsets: $2500$ subjects were selected to train the network, and two subsets of $1250$ each were used to validate and test the trained models \cite{bishop2006pattern, goodfellow2016deep}. 

As we stated before, pulse oximeter used in SHHS dataset provides a complementary quality signal. $\text{SpO}_2$ and HR were masked using this status signal, removing invalid data when the connection conditions were inadequate. Then, we linearly interpolated between the previous and posterior confident data. 

The signals were standardized before being used as input to the GRU. To standardize the input data to have zero mean and unit variance, the global mean and standard deviation were obtained using the train dataset and these values were used for all subsets: train, validation and test. The class imbalance for these subsets are $71.6\%$, $71.8\%$ and $71.5\%$, respectively.

\subsection{Recurrent neural networks: LSTM and GRU}
RNNs are a family of neural networks very useful for processing sequential data. In RNNs, sequences are processed one element at a time, storing in an ``internal state'' the information about all the history of the input. This persistence of information is achieved by internal loops (recurrent connections) that feedback the output \cite{lecun2015deep}. The classical RNNs are very difficult to train because the backpropagated gradients grow exponentially or vanish \cite{bengio1994learning}.

LSTM are networks especially designed to overcome the vanishing gradients, using a persistent internal state that can be modified by structures called gates \cite{hochreiter1997long}. For time step $t$ and cell $i$, the follow equations are performed:

\begin{equation}
\begin{aligned} \label{Eq:LSTM}
 \mathbf{f}_i^{(t)} &= \sigma (\mathbf{W}_{f, i} \cdot [\mathbf{h}_ i^{(t-1)}, \mathbf{x}^{(t)}] + \mathbf{b }_{f, i}),  \\ 
 \mathbf{g}_i^{(t)} &= \sigma (\mathbf{W}_{g, i} \cdot [\mathbf{h}_ i^{(t-1)}, \mathbf{x}^{(t)}] + \mathbf{b }_{g, i}),  \\ 
 \mathbf{\tilde{s}}_i^{(t)} &= \tanh (\mathbf{W}_{s, i} \cdot [\mathbf{h}_ i^{(t-1)}, \mathbf{x}^{(t)}] + \mathbf{b }_{s, i}),  \\ 
 \mathbf{s}_i^{(t)}  &= f_i^{(t)} s_i^{(t-1)} + g_i^{(t)} \tilde{s}_i^{(t)}, \\ 
 \mathbf{o}_i^{(t)} &= \sigma (\mathbf{W}_{o, i} \cdot [\mathbf{h}_ i^{(t-1)}, \mathbf{x}^{(t)}] + \mathbf{b }_{o, i}),  \\ 
 \mathbf{h}_ i^{(t)} &= \mathbf{o}_i^{(t)}  \tanh(\mathbf{s}_i^{(t)})
\end{aligned}
\end{equation}

\noindent where $\mathbf{f}_i^{(t)}$, $\mathbf{g}_i^{(t)}$ and $\mathbf{o}_i^{(t)}$ are the forget, input and output gates, respectively, $\mathbf{s}_i^{(t)}$ is the internal state, and $\mathbf{h}_ i^{(t)}$ is the output of the $i$-th LSTM cell. $\mathbf{W}_{(\cdot), i}$ and $\mathbf{b}_{(\cdot), i}$ are weights and bias, and $\sigma$ represents a sigmoid function \cite{goodfellow2016deep}. The input to the network is represented by $\mathbf{x}$.

The new internal state $\mathbf{s}_i^{(t)}$ depends on the last internal state (memory) and a ``filtered'' version of the last and the current inputs (update), controlled by forget and input gates. The output $\mathbf{h}_ i^{(t)}$ is a ``filtered'' version of the internal state, but multiplied by the output gate.

GRU is a simpler variation of LSTM that has become very popular \cite{cho2014learning}. The main difference with LSTM is that a single gate, called ``update'' gate, controls the forget and input of the internal state. The update equations are the following:

\begin{equation}
\begin{aligned} 
 \mathbf{u}_i^{(t)} &= \sigma (\mathbf{W}_{u, i} \cdot [\mathbf{h}_ i^{(t-1)}, \mathbf{x}^{(t)}] + \mathbf{b }_{u, i}),  \\ 
 \mathbf{r}_i^{(t)} &= \sigma (\mathbf{W}_{r, i} \cdot [\mathbf{h}_ i^{(t-1)}, \mathbf{x}^{(t)}] + \mathbf{b }_{r, i}),  \\ 
 \mathbf{\tilde{h}}_i^{(t)} &= \tanh (\mathbf{W} \cdot [\mathbf{r}_i^{(t)} \mathbf{h}_ i^{(t-1)}, \mathbf{x}^{(t)}] + \mathbf{b }_{s, i}),  \\ 
 \mathbf{h}_i^{(t)}  &= (1- \mathbf{u}_i^{(t)}) \mathbf{h}_i^{(t-1)} + \mathbf{u}_i^{(t)} \mathbf{\tilde{h}}_i^{(t)} 
\end{aligned}
\end{equation}

\noindent where $\mathbf{u}$ stands for ``update'' gate and $\mathbf{r}$ for ``reset'' gate. The update gate decides whether to copy (at one extreme of the sigmoid) or ignore (at the other extreme) the last state vector $\mathbf{h}_i^{(t-1)}$ by replacing with the new candidate of state vector $\mathbf{\tilde{h}}_i^{(t)}$. The reset gate controls which parts of the current state are used to compute the next state \cite{goodfellow2016deep}. The input to the GRU is represented by $\mathbf{x}$. The input shape, both to the LTSM and GRU, is a matrix which dimensions are sequence length, number of signals per batch and input size. In our work the input size is $2$, where the elements are HR and  $\text{SpO}_2$.

In this work, we prefered to use GRU instead of LSTM, since they have less memory usage. 

As it was presented until now, the RNN have a causal behavior, because the internal state stores only information from the past and present inputs. In sleep staging applications, we prefer that the classification depends on the entire input sequence. Having information from the past and the future allows a better understanding of the context and can eliminate ambiguities. Schuster and Paliwal \cite{schuster1997bidirectional} create a bidirectional RNN combining two RNN, one that moves forward through time and other that moves backward. 

Hereby, we evaluated bidirectional GRUs varying the number of cell units to achieve the best performance. The state vector $\mathbf{h}_i^{(t)}$ of each GRU was reinitialized to zeros at the beginning of each patient data. In this way, we avoid the spread of information from one patient to another.

\subsection{Softmax layer}

The GRU are responsible for learning the rules of transition and the temporal dynamics of the sequence. Then, a \emph{softmax layer} is used to classify each time step in two classes: awake and asleep. We apply a non-linear transformation by:

\begin{equation} \label{Eq:softLay}
  \mathbf{y} = \text{relu} (\mathbf{W} \mathbf{x} + \mathbf{b}) 
\end{equation}

\noindent where $\mathbf{W}$ and $\mathbf{b}$ are the weights and bias, respectively, and $\mathbf{x}$ is the input to this layer (that is, the output of the second bidirectional GRU). We use a rectified linear unit activation (i.e., $\text{relu}(x) = \max(0,x)$). This output vector $\mathbf{y}$ is mapped to a class probability with a softmax function. The used loss function and optimization algorithm are cross-entropy and mini-batch gradient-based optimization of stochastic objective functions called Adam \cite{kingma2014adam}, respectively.

\subsection{Classification resolution}

The designed algorithm yields a classification on a per second basis. This resolution is much higher than the one required by the AASM, which recommends labeling the sleep states every $30$ seconds. In order to adapt our algorithm to this standard, a majority vote within segments of $30$ seconds was conducted. The reported results correspond to this vote. Further, the targets with the sleep stage labels necessary to train the network have this ``resolution''. 

\section{Results} \label{sec:results}

We conducted several experiments to evaluate the network performance and the influence of its parameters. We describe the used performance measures and the selected optimizer. Then, we explain the training and testing stages of the network.

\subsection{Performance measures}

We compute several common statistics to evaluate the performance of the model: accuracy, sensitivity, specificity, precision, negative predictive value, and Cohen's Kappa coefficient \cite{sokolova2009systematic}. These measures were calculated individually for each patient and the averaged values are reported in table \ref{tab:Performance_measures}. In this work, we take the awake stage as a positive class. 

Further, we obtain two error measures for estimated TST. The average absolute error is defined as

\begin{equation} \label{Eq:err}
  \mathbf{E_1} = \frac{1}{N} \sum_{i=1}^{N} |\mathrm{TST}_i - \mathrm{T \hat{S}T}_ i |
\end{equation}

\noindent where $N$ is the total number of patients and $\mathrm{T \hat{S}T}_ i$ is the estimation of total sleep time, obtained by counting the segments classified as asleep. Similarly, the average absolute error percentage was calculated. It is defined as

\begin{equation} \label{Eq:err2}
  \mathbf{E_2} = \frac{1}{N}\sum_{i=1}^{N}  \frac{|\mathrm{TST}_i - \mathrm{T \hat{S}T}_ i |}{\mathrm{TST}_i} \cdot 100.
\end{equation}

\subsection{Network and optimizer parameters}

In order to evaluate the best hidden layer sizes in bidirectional GRUs we trained several models varying the number of cell units, taking the values $64$, $128$, and $256$. We trained models with preprocessed inputs. Two different alternatives were proposed as inputs: using only HR and using both HR and $\text{SpO}_2$. We tested all combinations of these parameters.


The parameters to Adam optimizer are learning rate $\alpha$, the exponential decay rate for the first moment estimates $\beta_1$, and the exponential decay rate for the second-moment estimates $\beta_2$ and they were set to $10^{-4}$, $0.9$ and $0.99$ respectively. The mini-batch size was set to $2$ due to memory restrictions of the GPU used for the training. 

\begin{figure}[t]
\centering
\includegraphics[width=0.3\textwidth]{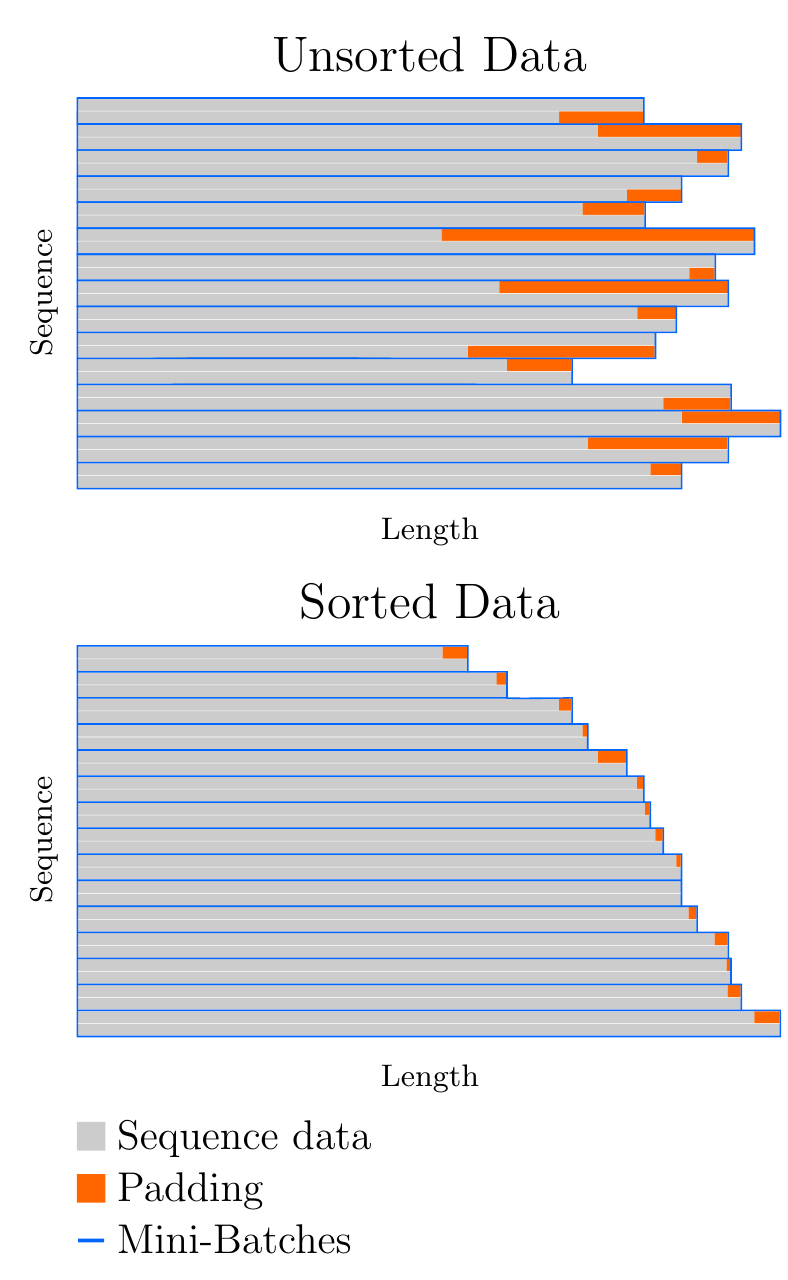}
\caption{Sorting the data by length avoid an excessive amount of padding, allowing faster data processing.}
\label{fig:padding_effect}
\end{figure}

In this application, the inputs are variable-sized sequences. RNN networks support input data with varying sequence lengths, but all the sequences in a mini-batch must be the same length to be packed. Therefore, we padded the sequences so that all the sequences in a mini-batch have the same length as the longest sequence in the mini-batch. To reduce the amount of padding, the input data was sorted by recording length \cite{goodfellow2016deep}. That is, the patients with similar-length recordings were grouped in the same batch. This effect is shown in Figure \ref{fig:padding_effect}. Logically, the network outputs corresponding to padded elements were not taking into account to calculate the loss function. The average length of the used data is $30432 \pm 2175$, while the maximum and minimum are $35970$ and $10800$ respectively.

\subsection{Network training}
\begin{figure}[t!]
\centering
\includegraphics[width=0.47\textwidth]{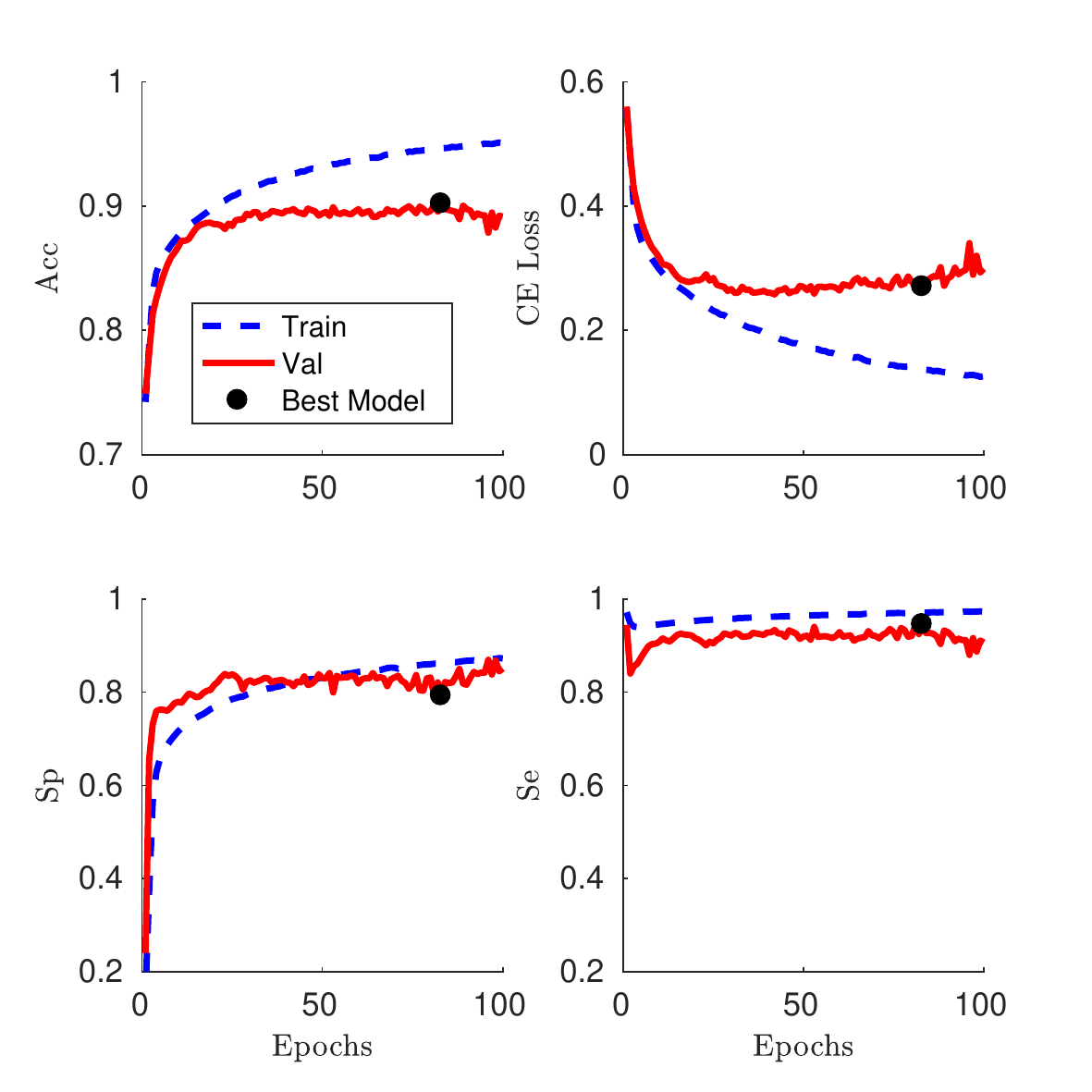}
\caption{GRU network training (256 hidden units) and model selection. Input: HR and $\text{SpO}_2$. The dashed (blue) and the solid (red) lines are the performances in train and validation dataset, respectively. The performances measures for this model are $90.13\%$ accuracy, $94.13\%$ sensitivity, and $80.26\%$ specificity.}
\label{fig:training_net_exp04c}
\end{figure}

We trained and selected the best model using the train and validation dataset. To do this, we adjusted the GRU weights iteratively using the training dataset in order to minimize the cross-entropy loss. After each epoch, the model was evaluated using the validation dataset. The number of epochs for the training was set to $100$. We used early stopping, selecting the model that had the best accuracy performance in the validation set. Figure \ref{fig:training_net_exp04c} shows the training process in one of the performed experiments and the selected model. In all the other experiments, the training had the same behavior. Then, the performance of the obtained optimum-model was evaluated in the test dataset \cite{bishop2006pattern, goodfellow2016deep}.

We also evaluated the use of $\mathbf{E_1}$ error as a cost function to optimize the network, since it is the most important measure for this application. However, the results obtained were not satisfactory.
 
\subsection{Performance in test dataset}
We evaluated the performance in an unseen test dataset of $1250$ patients. We have reported measures that are not affected by the imbalance between awake/asleep states, allowing to measure the model performance for each class: sensitivity, specificity, precision and negative predictive value \cite{fawcett2006introduction}. We have also calculated the Cohen's Kappa coefficient to measure inter-rater agreement between the predictions made by the algorithms and the ground-truth. Further, we have calculated the errors $\mathbf{E_1}$ and $\mathbf{E_2}$ to have a measure of error related to the application for which this work was intended.

Table \ref{tab:Performance_measures}\footnote{Abbreviations in Table \ref{tab:Performance_measures} and \ref{tab:others_performances}: $\text{Acc}$ for accuracy, $\text{Se}$ for sensitivity, $\text{Sp}$ for specificity, $\text{Prec}$ for precision, $\text{NPV}$ for negative predictive value, $\kappa$ for Cohen's Kappa coefficient, $\mathbf{E_1}$ for average absolute error, and $\mathbf{E_2}$ for average absolute error percentage.} shows the performances measures for train, validation and test datasets for all the tested networks. The performance improves slightly adding $\text{SpO}_2$ as input. The best result based on accuracy, loss function and Cohen's Kappa $\kappa$ coefficient is obtained using HR and $\text{SpO}_2$ and $2$ stacked layers of GRU with $256$ hidden units. Taking into account these performance measures, the results improve with the size of the hidden layer, as well as the overfitting risk.

Given that the network becomes bigger when the hidden layer size increases, and since we are also working with very long sequences (an $8$-hour record has $28800$ samples), the training also gets computationally very expensive. A trade-off exists between performance, the hidden layer size and the computational costs.

We performed a Friedman's test to assess if the results of the networks have significant differences. The $p$-value was $1.55 \times 10^{-131}$. This value suggests that at least one result is significantly different than others. Then, a multiple comparisons test was done to assess which pairs of results are significantly different. As a result of this test, it was obtained that the network that achieve the best result (HR and $\text{SpO}_2$ with $256$ hidden units) was significantly different from the others.
 

\begin{table*}[t]
\centering
\caption{Performance of the networks in train, validation and test datasets.}
\begin{tabular}{ c c c c c c c c c c c }
\hline \hline
  Inputs                & Network       &  Dataset 	    & $\text{Acc}$ 	   & $\text{Se}$ 	& $\text{Sp}$        & $\text{Prec}$ 	  & $\text{NPV}$ 	& $\kappa$	       & $\mathbf{E_1}$    & $\mathbf{E_2}$	    \\
\hline 	
                        &               & Train 	    & $94.31$          & $96.82$        & $85.74$ 	         & $94.76$            & $91.35$ 	    & $0.8396$         & $0.2193$          & $3.88\%$   	    \\
HR and $\text{SpO}_2$   & 2 GRU-(256)   & Validation    & $90.36$          & $94.60$        & $79.61$ 	         & $91.89$            & $85.53$	    	& $0.7414$	       & $0.4913$          & $9.30\%$    	    \\
                        &               & Test   	    & $\mathbf{90.13}$ & $94.13$        & $80.26$ 	         & $92.05$            & $\mathbf{84.68}$& $\mathbf{0.7400}$& $0.4842$          & $8.90\%$    	    \\
                        
                        &               &               &                  &                &                    &                    &                 &                   &                  &                   \\

                        &               & Train 	    & $92.56$          & $94.29$        & $86.36$ 	         & $94.80$            & $85.38$ 	    & $0.7975$          & $0.2687$         & $4.83\%$   	    \\
HR and $\text{SpO}_2$   & 2 GRU-(128)   & Validation    & $89.77$          & $92.49$        & $82.75$ 	         & $92.85$            & $81.66$	    	& $0.7339$	        & $0.4828$         & $8.97\%$    	    \\
                        &               & Test   	    & $89.69$          & $92.53$        & $82.58$ 	         & $\mathbf{92.78}$   & $81.58$ 	    & $0.7325$	        & $0.4745$         & $8.62\%$    	    \\     
                        
                        &               &               &                  &                &                    &                    &                 &                   &                &                      \\

                        &               & Train 	    & $91.04$          & $93.43$        & $84.54$ 	         & $93.68$            & $82.49$ 	    & $0.7607$          & $0.3536$         & $6.33\%$   	    \\
HR and $\text{SpO}_2$   & 2 GRU-(64)    & Validation    & $89.73$          & $92.65$        & $82.78$ 	         & $92.74$            & $81.14$	    	& $0.7321$	        & $0.4692$         & $8.70\%$    	    \\
                        &               & Test   	    & $89.59$          & $92.45$        & $\mathbf{83.12}$ 	 & $\mathbf{92.78}$   & $80.48$ 	    & $0.7306$	        & $\mathbf{0.4559}$& $\mathbf{8.28\%}$  \\
                        
                        &               &               &                  &                &                    &                    &                 &                   &                  &                    \\
                        
                        &               & Train 	    & $88.93$          & $92.02$        & $80.91$ 	         & $92.16$            & $79.55$ 	    & $0.7098$          & $0.4561$         & $8.43\%$   	    \\
HR                      & 2 GRU-(256)   & Validation    & $88.16$          & $91.75$        & $79.43$ 	         & $91.52$            & $78.78$	    	& $0.6911$	        & $0.5231$         & $9.85\%$    	    \\
                        &               & Test   	    & $87.99$          & $91.30$        & $79.99$ 	         & $91.62$            & $78.41$ 	    & $0.6905$	        & $0.5360$         & $9.68\%$    	    \\
                        
                        &               &               &                  &                &                    &                    &                 &                   &                  &                    \\
                        
                        &               & Train 	    & $90.67$          & $95.13$        & $78.76$ 	         & $91.76$            & $85.60$ 	    & $0.7446$          & $0.3989$         & $7.42\%$   	    \\
HR                      & 2 GRU-(128)   & Validation    & $89.20$          & $94.52$        & $76.51$ 	         & $90.58$            & $83.60$	    	& $0.7101$	        & $0.5078$         & $9.99\%$    	    \\
                        &               & Test   	    & $89.09$          & $\mathbf{94.23}$& $76.84$ 	         & $90.69$            & $83.57$ 	    & $0.7102$	        & $0.5220$         & $9.08\%$    	    \\
                        
                        &               &               &                  &                &                    &                    &                 &                   &                  &                    \\
                            
                        &               & Train 	    & $89.04$          & $93.69$        & $77.34$ 	         & $91.01$            & $82.13$ 	    & $0.7064$          & $0.4816$         & $9.18\%$   	    \\
HR                      & 2 GRU-(64)    & Validation    & $88.44$          & $93.54$        & $76.01$ 	         & $90.41$            & $81.61$	    	& $0.6908$	        & $0.5253$         & $10.26\%$    	    \\
                        &               & Test   	    & $88.17$          & $93.00$        & $76.58$ 	         & $90.55$            & $80.99$ 	    & $0.6882$	        & $0.5483$         & $10.12\%$    	    \\
\hline \hline
\end{tabular}
\label{tab:Performance_measures}
\end{table*}

\begin{figure}[t!]
\centering
\includegraphics[width=0.47\textwidth]{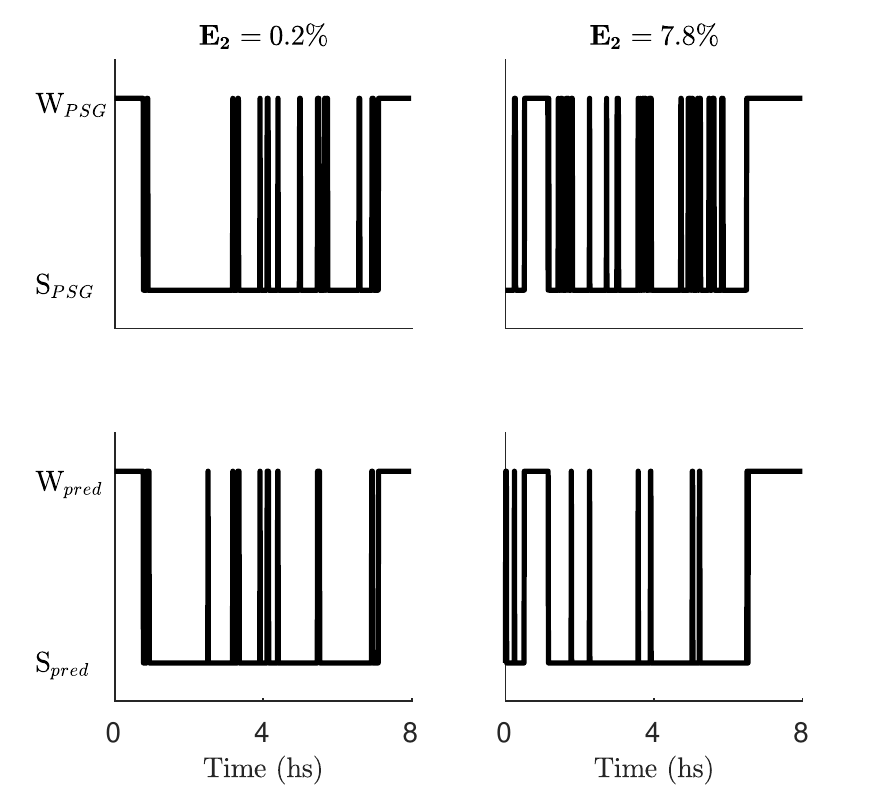}
\caption{An example of the best and average performing hypnograms using the network. The graphs on the left show the hypnogram obtained from the PSG (above) and the hypnogram obtained from the classifier (below) of the patient with the best relative error ($\mathbf{E_2}=0.2\%$). The graphs of the right show the hypnograms corresponding to a patient with an approximate average error ($\mathbf{E_2}=7.8\%$). The average absolute error $\mathbf{E_1}$ was $1$ and $25$ minutes, respectively.}
\label{fig:ejemplo_hpnograma}
\end{figure}

Figure \ref{fig:ejemplo_hpnograma} shows hypnograms for two of the participants in the database which have the best and approximate average error. 

\section{Discussion} \label{sec:discussion}

We developed a neural network based on bidirectional-GRUs to classify the sleep stage using signals provided by a pulse oximeter. The used inputs of the classifier were the raw signals, not hand-engineering extracted features as in classical machine learning approaches. Observing the obtained results, it can be seen that the incorporation of the $\text{SpO}_2$ signal produces a minor improvement in performance comparing to using only HR. $\text{SpO}_2$ signal can be useful for disambiguating confusing situations. For example, fast recoveries of the oxygen saturation after an apnea/hiponea event are usually associated with awakening events. Furthermore, decay of $\text{SpO}_2$ associated with apnea events can only appear during sleep.

The pulse oximeters present a great variability between different devices. Prior knowledge of the device plays a fundamental role in the interpretability of the results \cite{bohning2010comparability}. Despite this, many researches have achieved very good results using these signals with appropriate processing and it is becoming an important part of mobile and wearable devices.

The state-of-the-art results of automatic sleep staging are obtained using mostly EEG signals, but sometimes complementary information is extracted from other signals such as EOG, EMG and others. The regulation of HR contains information correlated with the sleep staging, but its interpretation is quite difficult. Further, the HR estimated by pulse oximetry has low temporal and frequencial resolution and it is strongly affected by motion artifacts. Despite this, through the use of information of history of the sequence, the designed network is able to achieve remarkable results. This simple model achieves results comparable with works that use more informative signals.

We compare our results with several previous works that performed automatic sleep classification. Since other signals are normally used in these works, comparisons can not be made directly. Furthermore, different databases and number of classes are used. In cases where it was necessary, the different sleep stages were considered as unique for comparative purposes. However, we will report which works discriminated sleep stages in more detail.

PPG signals and accelerometer were used in \cite{beattie2017estimation}. The authors considered $4$ sleep stages and tested the methods developed in $60$ normal sleepers subjects. In comparison with their work, we use only the HR calculated from the PPG, while they used the full PPG signal. Further, they have additional information from accelerometers. In spite of this, it can be seen that our algorithm obtains similar performance than their work for classification in asleep and awake. The performances obtained in \cite{beattie2017estimation} were $90.6\%$, $69.3\%$ and $94.6\%$ for accuracy, sensitivity and specificity, respectively. Although the accuracy and sensitivity obtained are very similar to those reported for our method, the specificity is significantly better in our work. Because the databases are unbalanced, it is important to observe the measures of specificity and sensitivity to be sure that the classification is not biased towards the majority class. 

PPG and HRV from PPG were used in \cite{uccar2016automatic} for awake/sleep classification using 10 patients. In their work, the performance obtained were $76\%$, $74\%$ and $80\%$ for accuracy, sensitivity and specificity, respectively. 

ECG signals were used in \cite{adnane2012sleep} to classify the sleep stage in awake or asleep. The used database  comprises $18$ patients. The performance values obtained were $80\%$, $69.1\%$ and $84.5\%$ for accuracy, sensitivity and specificity, respectively. 

ECG signals were also used in  \cite{yucelbacs2018automatic}, but in that work the sleep stages were classified in awake, REM and non-REM. Two different databases were used corresponding to $28$ patients in total. They reported discriminated performances for healthy subjects and patients. The accuracies for healthy subjects were $87.11\%$ and $77.02\%$ for the first and second database, respectively. For patients, the accuracies were $78.08\%$ and $76.79\%$. Notice that, the performance measures reported in their work are not the same as ours.

In our previous work \cite{casal2019sleep}, we extracted features from HR which were related to entropy and complexity measures, frequency domain and time-scale domain methods, and classical statistics. The best results were obtained by forward feature selection with SVM, in order to increase classification performance while reducing the feature space dimension. For the $30$-s length windows, performances achieved were  $73.7\%$, $54.6\%$ and $80.9\%$ of accuracy, sensitivity and specificity, respectively.

Finally, in \cite{malik2018sleep}, Malik et al. used the instantaneous heart rate (IHR) series obtained from ECG to classify wake/sleep status. They considered three different databases for validation, obtaining similar performances in all of them. In their private database, the accuracy, sensitivity and specificity were $83.1\%$, $52.4\%$, $89.4\%$, respectively. Further, they calculated the IHR from PPG and obtained similar performances. The architecture implemented in that work consist of five convolutional blocks, each one composed by two convolutional network. These blocks extract the features from the inputs. Then, these features are classified with a fully-connected network. In summary, the network have $12$ layers considering both convolutional and fully-connected layers.

Table \ref{tab:others_performances} summarizes these results for similar works. The PPG results from Malik et al. \cite{malik2018sleep} reported in this table correspond to training and testing using PPG. The authors performed several experiments to evaluate the transfer proficiency of the model among different monitoring devices that are not presented in the table. More related results can be found in \cite{yucelbacs2018automatic}.

\begin{table*}[t]
\centering
\caption{Comparison with the literature.}
\label{tab:others_performances}
\begin{tabular}{ c c c c c c c c  c c }
\hline \hline
Method                                      & Signal 		& Classes 	  &  Patients    & Epoch time	& $\text{Acc}$ 	& $\text{Se}$ & $\text{Sp}$	 & $\text{Prec}$ & $\text{NPV}$ \\
\hline 
\cellcolor{lg} Our work            & \cellcolor{lg} HR and $\text{SpO}_2$  & \cellcolor{lg} $2$ & \cellcolor{lg} $2500$ (train) & \cellcolor{lg} $30$ s & \cellcolor{lg} $94.31$  & \cellcolor{lg} $96.82$ &\cellcolor{lg}  $85.74$ & \cellcolor{lg} $94.76$ &\cellcolor{lg}  $91.35$     \\
\cellcolor{lg}  (256-biGRUs)       & \cellcolor{lg}                        & \cellcolor{lg} $2$ & \cellcolor{lg} $1250$ (val)   & \cellcolor{lg} $30$ s & \cellcolor{lg} $90.36$  & \cellcolor{lg} $94.60$ & \cellcolor{lg} $79.61$ & \cellcolor{lg} $91.89$ &\cellcolor{lg}  $85.53$     \\
\cellcolor{lg}                     & \cellcolor{lg}                        & \cellcolor{lg} $2$ & \cellcolor{lg} $1250$ (test)  & \cellcolor{lg} $30$ s & \cellcolor{lg} $90.13$  & \cellcolor{lg} $94.13$ & \cellcolor{lg} $80.26$ & \cellcolor{lg} $92.05$ &\cellcolor{lg}  $84.68$     \\
                                                &                       &         &      	           &            &          	    &          	  &          	 &          	&              	\\

Beattie et al. \cite{beattie2017estimation} 	& PPG + acc 	& $4$ 	  & $60$ 	           & $30$ s     & $90.6$ 	    & $69.3$ 	  & $94.6$ 	     & $70.5$ 	    & $94.3$ 	    \\

                                                &                       &         &      	           &            &          	    &          	  &          	 &          	&              	\\
                                                
                                                & ECG (CGMH-val) & $2$ 	  & $27$ 	           & $30$ s     & $83.1$ 		& $52.4$ 	  & $89.4$ 	     & $50.5$ 	    & $90.1$ 		    \\
Malik et al. \cite{malik2018sleep} 	    	    & ECG (DREAMS)          & $2$ 	  & $20$ 	           & $30$ s     & $81.4$ 		& $53.1$ 	  & $87.1$ 	     & $45.2$ 	    & $90.2$ 		    \\
                                                & ECG (UCDSADB)         & $2$ 	  & $25$ 	           & $30$ s     & $73.7$ 		& $43.4$ 	  & $81.9$ 	     & $39.2$ 	    & $84.3$ 		    \\
                                                & PPG (CGMH-val) & $2$     & $27$ 	           & $30$ s     & $84.2$ 		& $53.6$ 	  & $90.9$ 	     & $53.6$ 	    & $90.1$ 		    \\                                                
                                                
                                                &                       &         &      	           &            &          	    &          	  &          	 &          	&              	\\

                                                & PPG                   & $2$ 	  & $10$ 	           & $30$ s     & $76.8$ 	    & $76.0$ 	  & $77.0$ 		 & $41.2$ 	    & $93.8$  	    \\
Uçar et al. \cite{uccar2016automatic}		    & HRV		            & $2$ 	  & $10$ 	           & $30$ s     & $72.4$ 	    & $74.0$ 	  & $72.0$ 		 & $35.9$ 	    & $92.9$ 	    \\
                                                & PPG + HRV             & $2$ 	  & $10$ 	           & $30$ s     & $76.7$ 	    & $80.0$ 	  & $76.0$ 		 & $41.4$	    & $94.7$	    \\
						
                                                &                       &         &      	           &            &          	    &          	  &          	 &          	&              	\\
                                                
Adnane et al. \cite{adnane2012sleep} 	    	& ECG 			        & $2$ 	  & $18$ 	           & $30$ s     & $80.0$ 		& $69.1$ 	  & $84.5$ 	     & $64.5$ 	    & $87$ 		    \\

                                                &                       &         &      	           &            &          	    &          	  &          	 &          	&              	\\

Casal et al. \cite{casal2019sleep}               & HR                   & $2$ 	  & $4500$ 	           & $30$ s     & $73.7.9$ 		& $80.9$ 	  & $54.6$ 	    & $48.6$ 	    & $84.65$ 		\\
\hline \hline
\end{tabular}
\end{table*}

Due to the heterogeneity of the data and the experiments, it is not a simple task to compare the performances in the classification of sleep stages. Despite this, it can be said that the results obtained in this work are similar to those obtained using signals that are much informative and reliable, but also more difficult and expensive to be registered. Further, the size of the database used here is larger than those used in other works, reducing the risk of overfitting. A larger database allows to obtain a better generalization capability, especially with complex classifiers \cite{goodfellow2016deep}.



\section{Conclusions}
In this study, we designed an GRU-based model to classify the sleep stage in awake or asleep using HR and $\text{SpO}_2$ signals. It has been shown that relatively simple architectures can achieve good results in this field and that the information of HR is very useful to detect sleep. The $\text{SpO}_2$ allows a slight improvement in performance. As far as we know, the proposed network outperforms the state-of-the-art algorithms that used signals that are harder to acquire (except those with EEG). We have addressed a limitation of all apnea diagnosis methods based only on desaturation with a relative simple RNN. Further, the network can be easily adapted to other applications like drowsy driver monitoring, wearable devices for personal health monitoring, among others. The database used by us is much bigger than the ones used in related works. As future work, we will use these networks with other algorithms to detect apnea/hypopnea events with the aim of OSAHS diagnosis.

%
%

\bibliographystyle{elsarticle-num}
\bibliography{refs}

\end{document}